\documentclass{aa} 

\usepackage{aacite}
\usepackage{graphicx}

\newcommand{\Ha}{\mbox{${\rm H\alpha}$}}
\newcommand{\Hb}{\mbox{${\rm H\beta}$}}
\newcommand{\Hg}{\mbox{${\rm H\gamma}$}}
\newcommand{\Line}[3]{\Ion{#1}{#2}\,$\lambda$\,#3}
\newcommand{\Ion}[2]{#1{\,\scriptsize #2}}
\newcommand{\Mwd}{\mbox{$M_{\rm wd}$}}

\newcommand{\Rsec}{\mbox{$R_{\rm sec}$}}
\newcommand{\Msec}{\mbox{$M_{\rm sec}$}}
\newcommand{\Porb}{\mbox{$P_{\rm orb}$}}
\newcommand{\Msun}{\mbox{$M_{\odot}$}}
\newcommand{\ecsa}{\mbox{$\rm erg\;cm^{-2}s^{-1}\mbox{\AA}^{-1}$}}
\begin{document}
\thesaurus{06      
          (02.01.1 
           04.19.1 
           08.02.2 
           08.02.2 
           08.09.2 
           08.14.2 
          )}
\title{HS\,0907+1902: a new 4.2\,hr eclipsing dwarf nova}
\author{B.T. G\"ansicke \inst{1},
	R.E. Fried\inst{2},
        H.-J. Hagen\inst{3},
        K. Beuermann \inst{1},
        D. Engels\inst{3},
        F.V. Hessman\inst{1},
        D. Nogami\inst{1}
        \and
        K. Reinsch\inst{1}
}

\offprints{boris@uni-sw.gwdg.de}
\institute{ 
Universit\"ats-Sternwarte, Geismarlandstr. 11, D-37083 G\"ottingen, Germany
\and Braeside Observatory, PO Box 906, Flagstaff, AZ 86002, USA
\and Hamburger Sternwarte, Universit\"at Hamburg, Gojenbergsweg 112,
D-21029 Hamburg, Germany}
\date{Received \underline{\hskip2cm} ; accepted \underline{\hskip2cm} }
\authorrunning{B.T. G\"ansicke et al.}
\titlerunning{}
\maketitle
\begin{abstract}
We report on follow-up spectroscopy and photo\-metry of the cataclysmic
variable candidate HS\,0907+1902 selected from the Hamburg Quasar
Survey.
$B$, $V$, and $R$ photometry obtained during the first observed
outburst of HS\,0907+1902 ($V\approx13$) reveals deep eclipses 
($\Delta V\approx2.1-2.8$) and an orbital period $\Porb=4.2$\,h with
the eclipse depth decreasing to the red. The outburst eclipse profiles
are symmetric, indicating an axisymmetrical brightness distribution in
the accretion disc. We derive an inclination
$i\approx73^{\circ}-79^{\circ}$ from the eclipse duration.
The quiescent spectrum obtained with the Hobby-Eberly Telescope shows
double peaked emission lines of \Ion{H}{I}, \Ion{He}{I} and
\Ion{Fe}{I,II} and clearly identifies the system as a dwarf
nova. Absorption features of the secondary star are detected at red
wavelengths from which a spectral class M$3\pm1.5$ and a distance of
$d=320\pm100$\,pc are derived.

\end{abstract} 

\section{Introduction}
Cataclysmic variables (CVs) may be discovered by various
means. Historically, most of them were found because of their {\em
cataclysmic} nature, i.e. strong variability. This is especially valid
for dwarf novae, which show quasi-regular outbursts in the visual of
up to 8 magnitudes. With the advent of space-based X-ray telescopes, a
new class of CVs was discovered, containing magnetic white dwarfs as
accretors. The ROSAT and EUVE missions were extremely successful in
discovering this type of CVs (e.g. Beuermann
1998\nocite{beuermann98-1}).

However, a large number of CVs are neither prominent X-ray sources,
nor strongly variable. In non-magnetic CVs with a constantly high mass
transfer rate --~novalike variables~-- the accretion disc remains in a
perpetual hot state, turning them into unspectacular blue objects.
Similarly, dwarf novae with low outburst amplitudes or long outburst
cycles are likely to slip the attention of sky patrols.
As a result, the sample of known CVs \cite{downesetal97-1} is skewed
by selection effects, and the actual space density of CVs is a matter
of serious debate (e.g. de Kool 1992\nocite{dekool92-1} and Patterson
1998\nocite{patterson98-1}).

The Hamburg Schmidt objective prism survey (HQS, Hagen et
al. 1995\nocite{hagenetal95-1}), originally aimed at the detection of
a magnitude-limited sample of bright quasars (V=13--17.5), provides a
rich source of CV candidates selected because of their {\em
spectroscopic} properties.  Up to date, only few CVs have been
serendipitously identified from the HQS: HS\,0551+7241
\cite{dobrzyckaetal98-1}; HS\,1023+3900 \cite{reimersetal99-1}; and
HS\,1804+6753 (=EX\,Dra)
\cite{billingtonetal96-1,fiedleretal97-1}. The latter two objects show
the strength of the spectroscopic selection of CV candidates:
HS\,1023+3900 is a magnetic CV with a very low accretion rate and no
X-ray emission, and HS\,1804+6753 is a bright eclipsing dwarf nova
with low-amplitude outbursts, both stars were unlikely to be detected
with the ``classic'' selection mechanisms described above.

We have initiated a search for new CVs selected from the HQS with
follow-up observations of CV candidates detected also in the ROSAT
Bright Source catalogue \cite{vogesetal99-1}. They were identified as
possible CVs by Bade et al. \cite*{badeetal98-1} because of the Balmer
line emission seen in their HQS prism spectra. HS\,0907+1902
(\,=\,1RXS\,J090950.6+184956) was independently confirmed
spectroscopically as CV at the BAO (X. Jiang, private communication).
Here we report on the first photometric and spectroscopic results for
HS\,0907+1902.

\section{Observations and results}

\subsection{Photometry\label{s-phot}}
\begin{figure}[t]
\includegraphics[width=8.8cm]{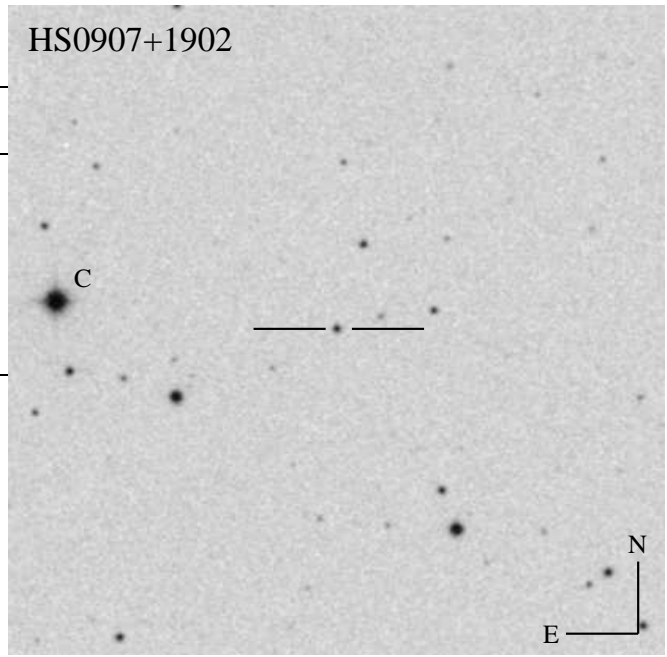}
\caption[]{\label{f-fc} Finding chart ($7\arcmin\times7\arcmin$) for
HS\,0907+1902 obtained from the Digitized Sky Survey. The coordinates of
the star are $\alpha(2000)= 09^h09^m50.5^s$ and
$\delta(2000)=+18^{\circ}49\arcmin47\arcsec$. The TYCHO comparison
star is marked 'C'}
\end{figure}

Several nights of differential photometry of HS\,0907+1902
(Fig.\,\ref{f-fc}) were obtained at Braeside Observatory, Arizona,
with a 41\,cm reflector equipped with a SITe\,512 CCD camera
(Table\,\ref{t-obs}). $B$ and $V$ magnitudes of HS\,0907+1902 were
derived relative to the $V=11.08$ and $B-V=0.86$ comparison star
(\,=\,Tycho2 1404-1852-1).

\begin{figure}[t]
\includegraphics[angle=270,width=8.8cm]{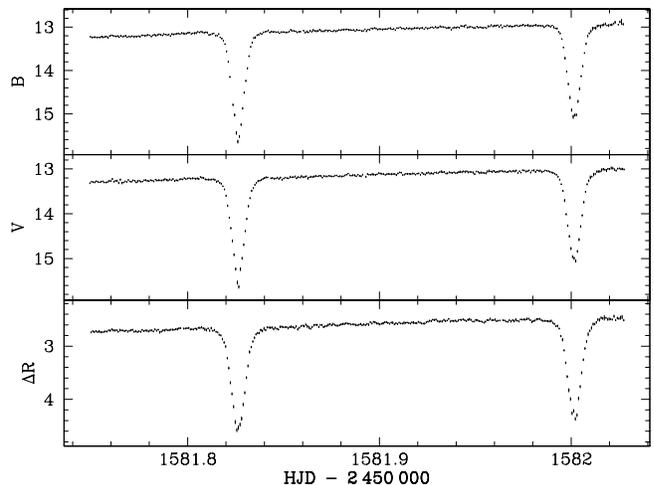}
\caption[]{\label{f-feb7} $B, V, R$ light curves obtained on February
7, 2000.}
\end{figure}

The first night of the observation run on 2000 February\,7 showed
HS\,0907+1902 at a magnitude of $V\sim13$. As Bade et
al. \cite*{badeetal98-1} estimated $B\approx16.4$ from the HQS prism
spectra, this clearly indicated that we had detected the first dwarf
nova outburst of HS\,0907+1902. The light curves (Fig.\,\ref{f-feb7})
show deep eclipses ($\Delta B=3.0$, $\Delta V=2.6$, and $\Delta
R=2.1$) with a period of 4.2\,h and low flickering activity. The mean
$V$ magnitude increased by $\sim0.2$ throughout the night, indicating
that HS\,0907+1902 was still on the rise to the maximum of the
outburst. No orbital modulation (hump) was detected.
The outburst was independently discovered by observers of the Variable
Star Network (VSNET) who reported the outburst of HS\,0907+1902 on
February\,11 (vsnet-alert
4176)\footnote{http://www.kusastro.kyoto-u.ac.jp/vsnet/Mail/vsnet-alert/msg04176.html}.

%

\begin{table}[t]
\caption[]{Observation log\label{t-obs}.}
\begin{flushleft}
\begin{tabular}{lcccc}
\hline\noalign{\smallskip}
HJD Start       & Int. time & \# of & Mode & Filter /   \\ 
(2\,450\,000+)  & [s]       & observations &      & Resolution \\
\hline\noalign{\smallskip}
1471.9798 & 600 & 2 & spect & 7$-$13\,\AA \\
1581.7493 & 58  & 330, 334, 333   & phot & $B, V, R$ \\
1585.6124 & 78  & 209, 209, 202   & phot & $B, V, R$ \\
1589.7248 & 50  & 390             & phot & white light \\
1590.6193 & 75  & 389             & phot & $V$ \\
1597.6189 & 35  & 64              & phot & white light \\
1599.6479 & 35  & 462             & phot & white light \\
\noalign{\smallskip}\hline
\end{tabular}
\end{flushleft}
\end{table}
%

The $B$, $V$, and $R$ light curves obtained on February\,11 cover one
eclipse and are very similar to those of February\,7, but the system
was somewhat fainter ($V\approx13.8$) and a decline by $\sim0.1$\,mag
is observed during the night. Hence, the maximum of the outburst
occurred between February\,7 and 11. HS\,0907+1902 was apparently in
outburst during the epoch of the plates of the Hubble Space Telescope
Guide Star Catalogue, which lists $V=12.53$. This value might be taken
as the brightest outburst magnitude so far recorded.

\begin{figure}[t]
\includegraphics[angle=270,width=8.8cm]{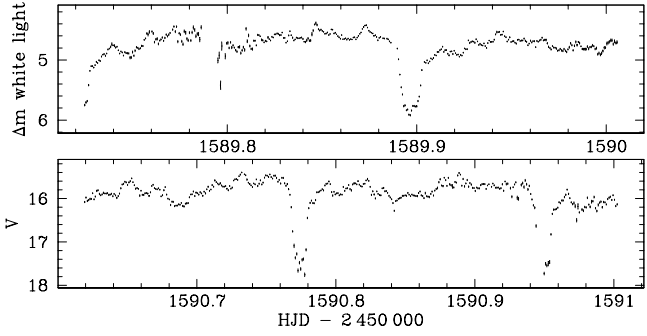}
\caption[]{\label{f-feb15} Top panel: White light curve obtained on
February 15. Bottom panel: $V$ light curve obtained on February 16.}
\end{figure}

On February\,15, the system appeared to be much fainter, and, due to
poor weather conditions, we decided to obtain filterless photometry to
maximise the time resolution. Flickering with an amplitude of
$\sim0.4$\,mag was apparent, typical of dwarf novae in
quiescence. Surprisingly, the light curve shows no strong orbital
modulation, which is normally the signature of a bright spot where the
accretion stream impacts the disc. One eclipse egress and one full
eclipse were covered.
On February\,16, we obtained a $V$ band light curve near $V\approx16$
(Fig.\,\ref{f-feb15}).  Finally, an eclipse egress and one complete
eclipse were covered on February\,23 and 25, respectively, again in
white light photometry, with the same out-of-eclipse magnitude as on
February\,15. HS\,0907+1902 was, therefore, already in quiescence on
the $15^\mathrm{th}$.

\smallskip\noindent 
{\em Ephemeris.} 
From the seven observed eclipses we derive  the following ephemeris:
\begin{equation}
\label{e-ephemeris}
\phi_0 = \mathrm{HJD}\,2451581.8263(1) +  0.175446(3)\times E
\end{equation}
where $\phi=0$ is defined as the mid-eclipse phase, equivalent to the
inferior conjunction of the secondary star. Errors in the last digit
are given in brackets. Table\,\ref{t-eclipses} lists the  eclipse timings.

\begin{figure}
\includegraphics[angle=270,width=8.8cm]{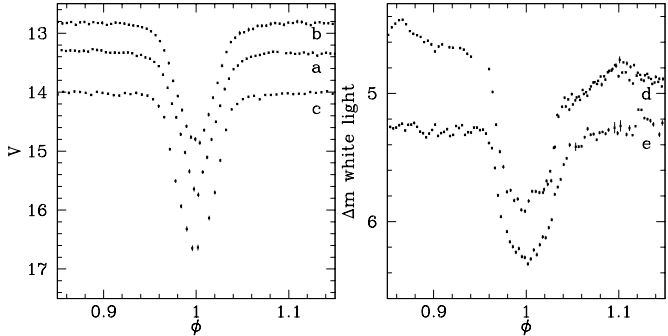}
\caption[]{\label{f-zoom} Left: $V$ band eclipse profiles obtained
during the outburst on February\,7 (a and b, b has been offset by
-0.5\,mag) and 11 (c), folded with the ephemeris
eq.\,(\ref{e-ephemeris}).  The light curves have been detrended from
the linear increase/decrease of the out-of-eclipse magnitude (see
Fig.\ref{f-feb7}). Right: white light eclipses obtained during
quiescence on February\,15 (d) and 25 (e, offset by +0.5\,mag).}
\end{figure}

\smallskip\noindent
{\em Eclipse shape.} 
We observed three eclipses during the dwarf nova outburst. In order to
compare the eclipse profiles, a linear fit was made to the
out-of-eclipse light curves and subsequently used to detrend the light
curves. The detrended eclipse profiles, folded with the above
ephemeris, are displayed in Fig.\,\ref{f-zoom}. The wings of the
eclipse profiles are perfectly symmetric with respect to $\phi=0.0$,
indicating an axisymmetrical brightness distribution in the accretion
disc in HS\,0907+1902.
The brightness at eclipse minimum is higher than that observed during
quiescence ($V\approx17.6$, Fig.\,\ref{f-feb15}), so the accretion
disc is not totally eclipsed. Interestingly, the centre of the eclipse
profile is variable in shape: while the first eclipse
(labelled 'a') on February\,7 has a round minimum, the second one
('b') has a flat bottom, which hints to an increase in
brightness of the non-eclipsed part of the accretion disc and could
explain the observed increase of the overall brightness of the
system. In the context of disc-outburst theory (e.g. Osaki
1996\nocite{osaki96-1}) this behaviour would be expected if the
observed outburst is of inside-out nature. The width of the eclipse at
half depth is $\Delta\phi_{1/2}\approx0.06$ during outburst, which is
somewhat lower than e.g. in IP\,Peg ($\Delta\phi_{1/2}\approx0.09$).
The measured eclipse duration implies
$i\approx73^{\circ}-79^{\circ}$ for a mass ratio $q=\Mwd/\Msec$ in
the range $1.25-3$.

\begin{table}[t]
\caption[]{Eclipse timings from the Braeside photometry\label{t-eclipses}.}
\begin{flushleft}
\begin{tabular}{rrrl}
\hline\noalign{\smallskip}
\multicolumn{1}{c}{HJD$(\phi_0)$}  &
\multicolumn{1}{c}{O--C} & 
\multicolumn{1}{c}{Filter} & 
\multicolumn{1}{c}{State}  \\
\multicolumn{1}{c}{(2\,450\,000+)} & \Porb & & \\
\hline\noalign{\smallskip}
1581.8263 & $ 2.5\times10^{-4}$ & $B,V,R$ & outburst \\
1582.0017 & $-4.8\times10^{-4}$ & $B,V,R$ & outburst \\
1585.6861 & $-1.3\times10^{-4}$ & $B,V,R$ & outburst \\
1589.8969 & $ 2.4\times10^{-4}$ & white light  & quiescence \\
1590.7742 & $ 7.2\times10^{-4}$ & $V$     & quiescence \\
1590.9496 & $ 4.9\times10^{-4}$ & $V$     & quiescence \\
1599.7219 & $ 3.9\times10^{-4}$ & white light  & quiescence \\
\noalign{\smallskip}\hline
\end{tabular}
\end{flushleft}
\end{table}

Unfortunately, during quiescence only two full eclipses were covered
with data of satisfactory quality (Fig.\,\ref{f-zoom}), which leaves the
following statements somewhat speculative.
Taken at face value, the eclipse observed on February\,15 broadly
resembles that of Z\,Cha during quiescence
\cite{cook+warner84-1}: the eclipse minimum is followed by a steep jump
in brightness, which is followed by a smooth egress to the
out-of-eclipse level. In Z\,Cha, the final egress is due to the
appearance of the bright spot, whereas the sudden jump from eclipse
minimum corresponds to the egress of the white dwarf. The hypothetical
white dwarf egress in HS\,0907+1902 would result in a relatively short
eclipse of the primary ($\Delta\phi\sim0.04$), which agrees with the
conclusions on $i$ and $q$ obtained above from the high state eclipse
profiles. However, the eclipse obtained on February\,25 has a less
structured shape, and higher S/N data are needed to decisively derive
the binary parameters.

\begin{figure*}[t]
\includegraphics[angle=270,width=18cm]{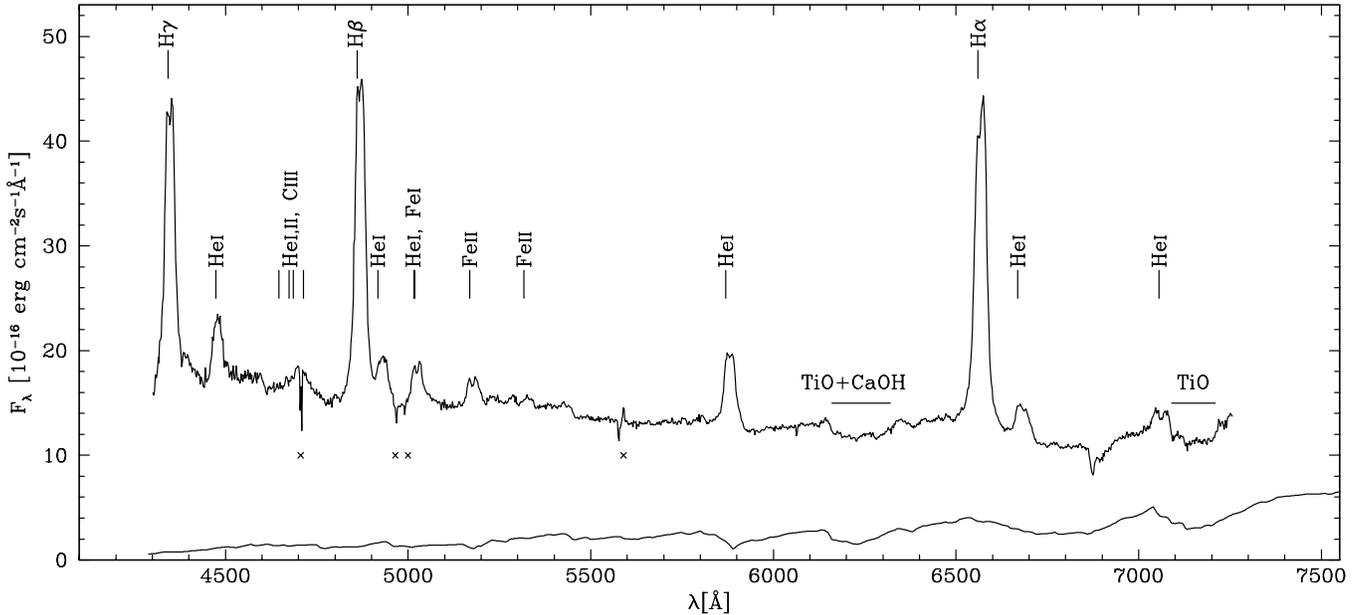}
\caption[]{\label{f-spectrum} HET/LRS spectrum of HS\,0907+1902 in
quiescence. Major emission lines are identified. Also shown is the
spectrum of the M3 dwarf Gl352ab, which has been scaled to $V=18.13$ in
order to match the observed TiO and CaOH absorption observed in
HS\,0907+1902.
Crosses mark structures due to CCD defects and poor sky line subtraction.}
\end{figure*}

\subsection{Spectroscopy\label{s-spect}}
On 1999 October 20, two identification spectra of HS\,0907+1902 were
obtained with the low resolution spectrograph (LRS,
$\lambda/\Delta\lambda\approx520$) on the 9.2\,m Hobby-Eberly
telescope (HET) on Mt. Fowlkes, Texas. An absolute flux calibration
of these spectra was not possible due to thermal drift in the alignment
of the 91 individual segments of the primary mirror.  We, therefore,
used the flux standard LDS749b to derive the instrumental response
function and adjusted the spectra of HS\,0907+1902 to the observed $V$
magnitude. One of the HET spectra is shown in Fig.\,\ref{f-spectrum}.

The spectrum of HS\,0907+1902 is typical of a dwarf nova in
quiescence, containing strong Balmer emission lines and weaker
\Ion{He}{I} and \Ion{Fe}{II} lines. There is possibly some weak
emission of \Line{He}{II}{4686}, blended with \Line{He}{I}{4713} and
with the \Ion{C}{III}/\Ion{N}{III}\,$\lambda\lambda4640-4650$
complex. The double-peaked shape of the emission lines
is typical for high-inclination dwarf novae.
The equivalent widths of the most prominent lines are
\Ha\,=\,90\,\AA, \Hb\,=\,78\,\AA, \Hg\,=\,46\,\AA,
\Line{He}{I}{5870}\,=\,24\,\AA, \Line{He}{I}{4474}\,=\,10\,\AA. The 
FWHM, corrected for the instrumental resolution, are
\Ha\,=\,32\,\AA, \Hb\,=\,28\,\AA, \Hg\,=\,26\,\AA,
\Line{He}{I}{5870}\,=\,33\,\AA, \Line{He}{I}{4474}\,=\,26\,\AA.

The red end of the HET spectrum of HS\,0907+1902 shows signatures of a
late-type secondary star, namely the broad absorption blends of
TiO/CaOH ($\lambda\lambda6160-6320$) and of TiO
($\lambda\lambda7190-7210$). The flux increases for
$\lambda>7210$\,\AA, as expected for the contribution of a late type
star. Using a library of observed M-dwarf spectra, we derive a
spectral type of M$3\pm1.5$ for the secondary star in HS\,0907+1902.
This estimate agrees well with the observed spectral types of
secondaries in CVs with similar \Porb\ \cite{beuermannetal98-1}.
Fig.\,\ref{f-spectrum} shows the M3-dwarf Gl352ab scaled according to
the depth of the observed absorption features.

From the adjusted M-star spectra of stars with the above range of
spectral types, we measure an observed surface brightness of the flux
difference in the $\lambda\lambda7500/7165$\,\AA\ band of
$f_\mathrm{TiO}=(3.9\pm1.7)\times10^{-16}$\,\ecsa. From
Roche geometry and from Patterson's \cite*{patterson84-1} mass-radius
relation for main-sequence stars we estimate
that $\Msec=0.42\Msun$ and $\Rsec=3.17\times10^{10}$\,cm. Finally,
applying the calibration for the $F_\mathrm{TiO}$ surface brightness
of late-type stars of Beuermann \& Weichold
\cite*{beuermann+weichhold99-1}, we obtain a distance of
$d=320\pm100$\,pc, corresponding to a distance modulus of
$m-M=7.5\pm0.7$.

If we assume an outburst magnitude of $V=12.5$ (Sect.\,\ref{s-phot}),
we derive an absolute magnitude in outburst of $M_V=3.3$, where we
applied a correction $\Delta M_V(i)=1.6$ for an assumed inclination of
$80^{\circ}$.  Warner's \cite*{warner87-1} $M_V-\Porb$ relation gives
$M_V=4.6$ for $\Porb=4.2$\,h; this can be taken as a hint that the
true distance is on the lower side of our error range and that the
spectral type of the secondary is rather $\sim M4$.

\section{Conclusion}
We have discovered a bright new eclipsing dwarf nova with an orbital
period of 4.2\,h. Eclipsing CVs offer the best means of deriving
the system parameters such as stellar masses,  mass transfer
rates, and the structure of the accretion disc. With its long orbital
period, HS\,0907+1902 is only the fourth deeply eclipsing dwarf nova
above the $2-3$\,h period gap, the other ones being IP\,Peg,
HS\,1804+6753 (=EX\,Dra), and BD\,Pav. With a quiescent and an
outburst magnitude of $V\approx16$ and $V\approx12.5$, respectively,
HS\,0907+1902 is well suited for detailed follow-up studies.

\acknowledgements{BTG and DN were supported by DLR/BMBF grant
50\,OR\,9903\,6. The HQS was supported by the Deutsche
Forschungsgemeinschaft through grants Re 353/11 and Re
353/22. Braeside Observatory acknowledges the support of The Research
Corporation, The National Science Foundation (AST-92-180002), and the
Fund of Astrophysical Research.
The Hobby-Eberly Telescope is operated by McDonald Observatory on
behalf of The University of Texas at Austin, the Pennsylvania State
University, Stanford University, Ludwig-Maximilians-Universit\"at
M\"unchen, and Georg-August-Universit\"at G\"ottingen. The Marcario
Low Resolution Spectrograph is a joint project of the Hobby-Eberly
Telescope partnership and the Instituto de Astronom{\'\i}a de la
Universidad Nacional Autonoma de M{\'e}xico. We thank the referee
Dr. Osaki for  helpful comments.}

\end{document}